\documentclass[%
superscriptaddress,
reprint,
preprintnumbers,
 amsmath,amssymb,
 aps, physrev,
]{revtex4-2}

\usepackage{graphicx}
\usepackage{dcolumn}
\usepackage{bm}

    \usepackage{cancel, amsmath}

\usepackage{tikz}
\usetikzlibrary{shapes,positioning}

\newcommand{\red}{\textcolor{red}}

\usepackage{ulem}
\begin{document}


\title{\textbf{Quantifying Coupled Dynamics in 
Phase-Space from State Distribution Snapshots} 
}%

\author{Erez Aghion}
 \affiliation{Department of Physics, University of Louisiana at Lafayette, LA, 70504, USA}
\author{Nava Leibovich}%
 \email{Contact author: nava.leibovich@nrc-cnrc.gc.ca}
\affiliation{%
 National Research Council of Canada, NRC-Fields Mathematical Sciences Collaboration Centre, 
Toronto, Ontario, M5T 3J1, Canada.
}%




\begin{abstract}
We quantify nonlinear interactions between coupled complex processes, when the system is subject to noise and not all its components are measurable. Our method is applicable even when the system cannot be continuously monitored over time, but is rather observed  only in snapshots. Having only partial information about the local topology of the network and  observations of relevant interacting variables is sufficient to translate qualitative knowledge of interactions into a quantitative characterization of the coupled dynamics.   
   This approach turns a globally intractable problem into a sequence of solvable inference problems, to quantify complex interaction networks from incomplete snapshots of their  statistical state.  
\end{abstract}

\maketitle


\section{\label{sec:introduction}Introduction}

Quantifying interactions  between different components of a complex network, from stationary snapshots of its statistical state, is a long-standing challenge. 
For example, understanding cellular processes requires quantifying biochemical reaction rates between molecules, but in experiments,  data may be  gathered from static molecular concentrations 
 within a collection of individual cells \cite{banani2017biomolecular, svensson2018exponential, robinson2015flow, requena2023inferring}  or by analyzing omics data from multiple patients \cite{subramanian2020multi, dibaeinia2025interpretable}.
Another example is assessing interspecies interactions within complex ecological networks, which may rely on data collected from static states \cite{
kurtz2015sparse,dormann2017identifying, keane2011encounter}
 . Complex interactions, however, occur through dynamical processes, involving many different components in the network. 

Existing techniques for inferring interactions in coupled complex systems often use statistical tools  based on 
time series measurements \cite{warne2019simulation, guillen2013identification, casadiego2017model,klett2021non,seckler2022bayesian, kumar2025diffmap}. Methods subjected to specific interaction models and constraints \cite{timme2007revealing, brunton2016discovering, timme2014revealing, weinreb2018fundamental}, or discrete data \cite{wittenstein2022quantifying, leibovich2025determining} were also previously proposed. However, empirical data does not necessarily comply with such requirements. In this paper, we develop a method that does not necessitate temporal information,  complete observations of all the components within the system, or a pre-selected set of candidate functions for the interaction mechanisms. Importantly, our method is optimized for noisy dynamics and data subjected to measurement noise. Our proof-of-principle examples illustrate how local network measurements can be utilized for quantitative inference of coupled stochastic dynamics from observed data, using only static snapshots of naturally occurring population variability.  

Our work is motivated in part by  experiments that use
fluorescence microscopy to visualize single-cell and single-molecule dynamics of gene expression. For a review of existing inference methods for predictive models of stochastic gene expression, see~\cite{li2011central, buccitelli2020mrnas, aguilera2025methods}. Based on these models, we concentrate on systems evolving via $N$ coupled overdamped, Markovian Langevin equations \cite{gardiner1985handbook} for the state vector $\vec{x}=(x_1, \dots, x_N )$: 
\begin{align} 
   \partial _t  \vec{x}(t)= \vec{\mu}[\vec{x}(t)]  + \vec{\xi}(t). 
    \label{EqLangevins}
\end{align}
Here, each term in $\vec{\mu}$; $\mu_i(\vec{x})$,  is a force applied on a single variable, $x_i$ (where  $i\in\{1,\dots, N\}$), that can depend on a nonlinear combination of the parameters in $\vec{x}$. The meaning and units of $x_0,x_1...$ is determined by the system under consideration. Each term in $\vec{\xi}$; $\xi_i $, is Gaussian white noise with $\langle\xi_i \rangle =0$, $\langle\xi_i(t)\xi_i(t')\rangle=2D_{i}\delta(t-t')$ and $\langle\xi_i(t)\xi_j(t')\rangle=0$ for $i\neq j$, the amplitude of the noise is $2D_{i}\geq0$. Equation \eqref{EqLangevins} means that the dynamics of $x_i$ is coupled to the other variables through $\mu_i(\vec{x})$. 
\begin{figure}
    \centering
    \includegraphics[width= \columnwidth]{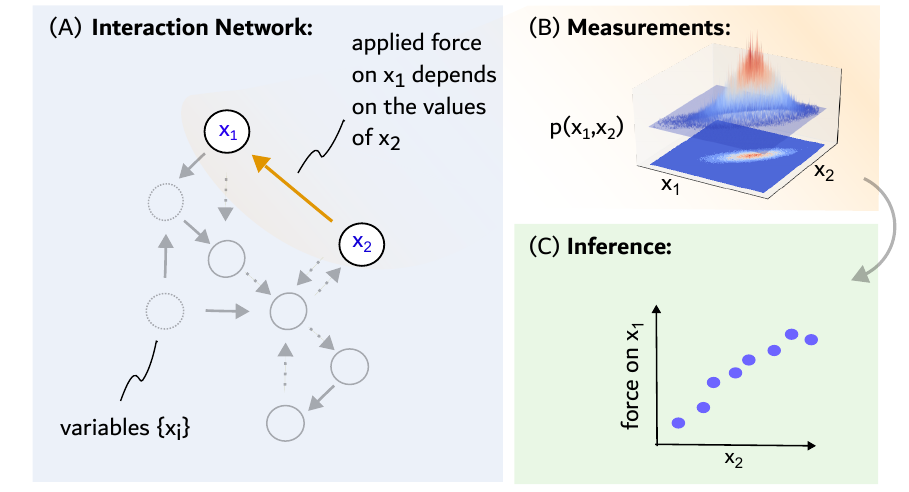}
    \caption{A scheme of the research goal. Using a sampled stationary probability density function to infer the applied force. (A) In an interaction network, solid (dotted) arrows indicate the effect of some variables on others, when it is experimentally accessible (inaccessible).  Our goal is to quantify the effect of $x_2$ on $x_1$. (B) The data provided for the analysis is a stationary snapshot of the mutual probability density function, $p(x_1,x_2)$, here illustrated.  (C) From the shape of $p(x_1,x_2)$ we devise a method to infer the coupling force on $x_1$ as a function of $x_2$.}
    \label{fig:scheme}
\end{figure}

The joint probability density function, $p(\vec{x},t)$, of the state vector $\vec{x}$ follows  Fokker Planck equation~\cite{gardiner1985handbook}
\begin{eqnarray}
    \frac{\partial p(\vec{x},t)}{\partial t} =  -\sum_{i=1}^N \frac{\partial}{\partial x_i} \mu_i(\vec{x}) \, p(\vec{x},t) + \sum_{i=1}^{N} \frac{D_{i}}{2}  \frac{\partial^2}{\partial x_i^2}    p(\vec{x},t).  \nonumber
 \\ 
 \label{EqFokkerPlanck}
\end{eqnarray}  
Given the force terms $\{\mu_i(\vec{x})\}_{i=1}^N$, Eq.~\eqref{EqFokkerPlanck} enables to predict the shape of $p(\vec{x},t)$ at any time $t$. However, here we focus on the inverse problem:  
Can we quantify $\mu_i(\vec{x})$ from stationary measurements of the probability density?

Stationary measurements of the multidimensional probability density function are obtained in experiments through various methods tailored to the specific variables under study. In one example, fluorescence  labeling and spectral imaging were used to quantify multi-species abundance and interactions within microbial communities \cite{valm2011systems,mickalide2019higher}. In a different example, to quantify mRNA-protein joint abundances in \cite{lin2019ultra}, the authors use an automated microfluidic system combined with an ultra-sensitive digital proximity ligation assay (dPLA). A review paper;  \cite{buccitelli2020mrnas}, describes a variety of multidimensional probability density function imaging experiments, allowing the exploration of coupled intercellular processes in regulatory mechanisms underlying cellular behavior.

\section{Main Results}
\label{SecMainResults}
 
 This manuscript focuses on systems with nonlinear interactions, coupling multiple stochastic variables. Here, the force applied on the $i$-th variable, $\mu_i(\vec{x})$, can depend on the state of all the other variables, $\vec{x}$, or a portion of them.  This system can be represented by a network with nodes, $\{x_i\}$, where the edges represent the interdependence between the variables,  Fig.~\ref{fig:scheme}.

Without loss of generality, we consider the inference of the applied force on the arbitrarily chosen variable; $x_1$. This force depends on $x_1$ and on the state of other variables $\vec{E}_1^- = (x_2, x_3,\ldots)$. Using network terminology, $\vec{E}_1^-$ are the variables with directed incoming edges (indicated by the superscript $(\cdot)^-$) to node $x_1$ (as indicated by the subscript,  $(\cdot)_1$). 
Our goal is to infer $\mu_1(x_1, \vec{E}_1^-)$ from the empirical observations of the probability density function, $p(x_1, \vec{E}_1^-)$.

Consider systems that can reach a steady state at the long-time limit, then the stationary Fokker-Planck equation projected on $x_1$ yields (see Sec.~\ref{SecMethod} for the derivation)
\begin{eqnarray}
    \frac{\partial}{\partial {x_1}}[\ln p(x_1)] = 2 \langle \mu_1(\vec{x}) | x_1 \rangle /D_1, 
    \label{EqMain1}
\end{eqnarray}
where $p(x_1)$ denotes the marginal steady-state probability density of $x_1$. The angular bracket notation $\langle \cdot |x_1\rangle$ means a conditional average for a given $x_1$. When the self regulation term in the force on the variable $x_1$ can be decoupled from the other variables as  $\mu_1(x_1, \vec{E}_1^-) = \mu_{1E}(\vec{E}_1^-) - \mu_{11}(x_1)$,  Eq.~\eqref{EqMain1} is re-written as
\begin{eqnarray}
\label{EqMain2}
    \frac{D_1}{2}\frac{\partial}{\partial {x_1}}[\ln p(x_1)] + \mu_{11}(x_1)    &=& \int d\vec{x} \, p(\vec{x}| x_1) \mu_{1E} ( \vec{x})  \\ 
    \nonumber
    &=& \int d\vec{E}_1^- \, p(\vec{E}_1^- | x_1) \mu_{1E} ( \vec{E}_1^-). 
\end{eqnarray}
Here, the integration takes place over the entire phase space. The second transition in Eq.~\eqref{EqMain2} is permissible because $\mu_{1E}$ depends solely on the variables within $\vec{E}_1^-$, allowing to average out the other variables. Replacing the integration with summation, we find that 
{\small \begin{eqnarray}
    \frac{\partial}{\partial {x_1}}\frac{D_1\ln p(x_1)}{2} + \mu_{11}(x_1) 
    =  \sum_{\substack{\vec{E}_1^-\\ \text{values}}} \Delta_{{\vec{E}_1^-}}p(\vec{E}_1^- | x_1) \mu_{1, E} (\vec{E}_1^-),
  \nonumber \\   \label{EqMainResults5}
\end{eqnarray}
}where $\Delta_{{\vec{E}_1^-}}$  are small constants, see Sec.~\ref{SecMethod}.

Empirical data inherently consists of a finite set of discrete data points. This means that the probability density function is empirically estimated at a limited set of values \( \left(x_1^i, ({{\vec{E}_1^-}})^j\right) \in \{x_1^i\} \times \{ \vec{E}_1^- \text{ values}\} \). 
To quantify $\mu_{1,E}$ means to identify the set of values \( \mu_{1,E}\left((\vec{E}_i^-)^j\right) \) for each \( (\vec{E}_i^-)^j \in \{ \vec{E}_1^- \text{ values}\} \) that provides the best agreement with Eq.~\eqref{EqMainResults5}  corresponding to the observed $p\left(x_1^i, ({{\vec{E}_1^-}})^j\right)$. This is achieved by leveraging a numerical optimization approach \cite{martins2021engineering}, 
see Sec.~\ref{SecMethod} and App.~\ref{AppA} and~\ref{App:ThreeComponentsAndMore}. 

In the following, we demonstrate the applicability of our method through multiple examples featuring various characteristic dynamics. These examples highlight several key aspects of the inference method. The applied force $\mu_{1,E}$ is embedded within the observed distribution $p(x_1,\vec{E}_1^-)$, where the latter is subject to sampling noise inherent in the observation process. Notably, we do not impose any specific functional form on $\mu_{1,E}$; instead, we find values for $\mu_{1,E}\left((\vec{E}_1^-)^j\right)$ for each $(\vec{E}_1^-)^j \in \{\vec{E}_1^- {\rm \ values}\}$. Achieving this objective requires neither a comprehensive understanding of the interaction mechanisms, nor the complete network topology, nor the observation of all variables within the system.

\subsection{Proof of Principle Examples}
\label{SecExamples}
We first illustrate our inference method on a biologically realistic model describing gene expression, building upon the \textit{central dogma of molecular biology}~\cite{crick1970central, chen1999modeling, li2011central, ingalls2013mathematical, aguilera2025methods}, see details in App.~\ref{App:ProofOfConceptExamples}. Second, we use three prototypical models with more complex network topologies. These models represent a selection of interaction motifs leading to: (A) bistable dynamics, (B) noise-controlled and (C) oscillatory dynamics, App.~\ref{App:ProofOfConceptExamples}. Third, we resolve two examples of interactions of in high-dimensional systems, App.~\ref{App:ProofOfConceptExamples}.  
\subsubsection{{A Model of Gene Expression}}
As our first example, we quantify the coupled dynamics of a synthetic, yet biologically realistic~\cite{aguilera2025methods, ingalls2013mathematical}, model of a gene expression process within the cell. Specifically, this model encompasses the gene, mature nuclear mRNA, cytoplasmic mRNA, proteins, and the interactions between these components, see detailed in App~\ref{App:ProofOfConceptExamples}.  

{To quantify the protein production rate as a function of  the number of cytoplasmic mRNA molecules experimentally, one can analyze single-cell fluorescence microscopy images. In this case,  single-molecule fluorescence in situ hybridization (smFISH) can be used for empirically quantifying the mRNA \cite{femino1998visualization}. Nascent chain tracking (NCT) can also be employed  to record snapshots of protein and mRNA signals \cite{morisaki2016real}. After these snapshots are taken, one needs to execute image processing techniques for extracting the relevant information from images - for example, to detect and count the mRNA molecules, and  differentiate between nuclear and cytoplasmic mRNA.  This involves implementing a computational method to segment the cytosol and nucleus for each cell in the simulated images, followed by counting the cytosolic mRNA and proteins in each cell, as   illustrated in Fig.~\ref{fig:BiologicalRealisticModel}(A).  To generate this image from our simulation results, we used a script provided in \cite{aguilera2025methods} for analyzing single-cell fluorescence microscopy images, see App.~\ref{App:ProofOfConceptExamples}. } 
 
 {Our method for quantifying the coupled dynamics represents an advancement in the analysis of information derived from images such as Fig.~\ref{fig:BiologicalRealisticModel}(A). By investigating the joint  probability density function of cytosolic mRNA and protein copy numbers, we accurately quantify the protein production rate relative to the levels of cytosolic mRNA molecules, as shown in Fig.~\ref{fig:BiologicalRealisticModel}(B).
 Importantly, at no stage in our analysis do we make any assumption about the functional shape of the  production rates, and we do not use any information from the other molecular concentrations within the cell, such as information about the concentration of $G$ and $R_n$.    

\begin{figure}
    \centering
    \includegraphics[width=1.0\linewidth]{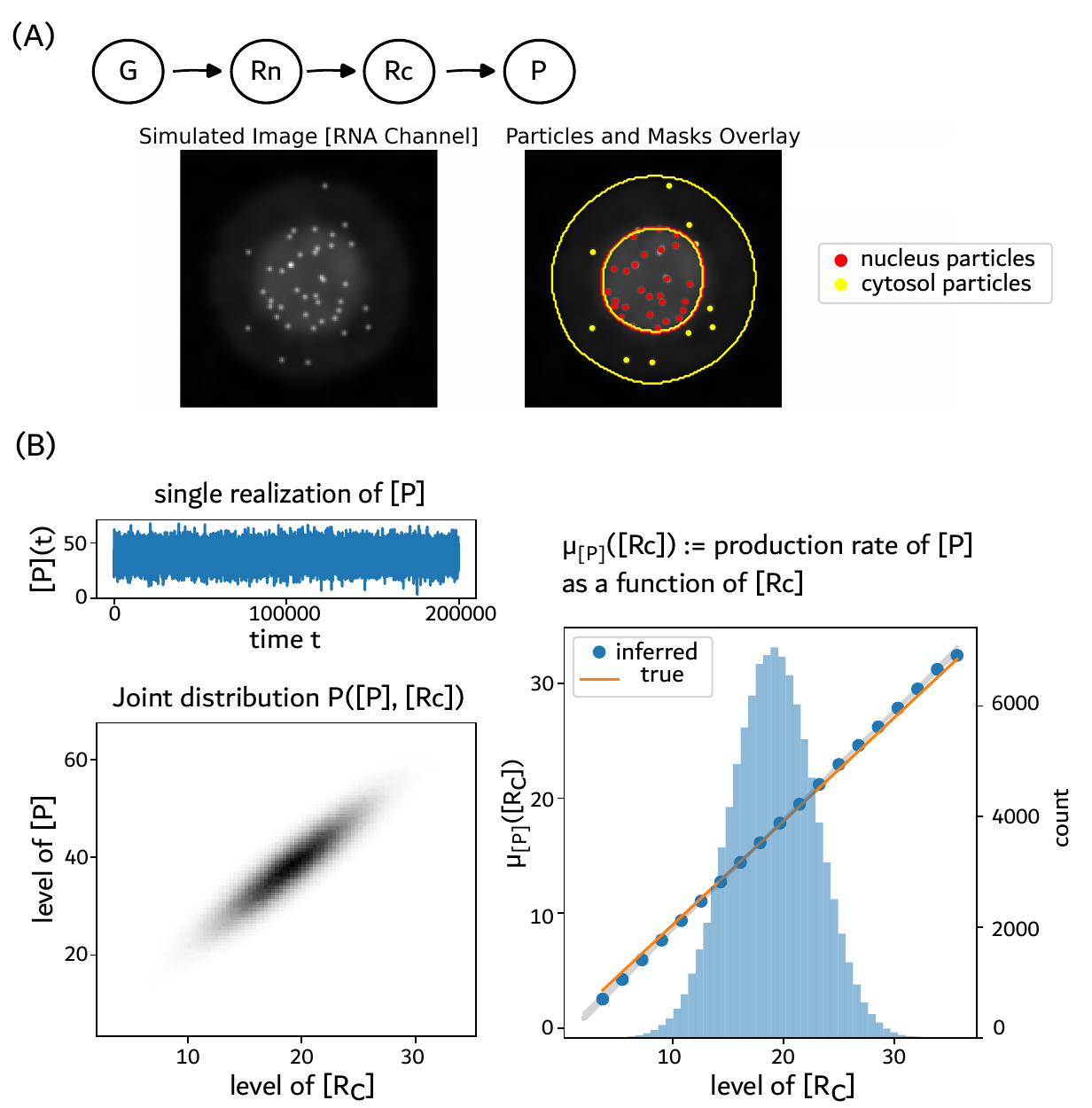}
  \caption{Simulation and analysis results for the Gene Expression model. Panel (A): This minimal model includes the components of gene [G], mature nuclear mRNA [Rn], cytoplasmic mRNA [Rc], and protein [P]. Arrows represent the direction of the transcription-translation process. Two representative images illustrate cell segmentation and particle detection in the RNA channel, see App.~\ref{App:ProofOfConceptExamples} for the simulation details. The left image shows the original simulated RNA channel image after Gaussian smoothing. On the right image, we added the overlays and segmentation masks used to distinguish between the nuclear participles (${\rm R_n}$) and cytosol particles (${\rm R_c}$), see \cite{aguilera2025methods}. The inner (outer) contoured area delineates the segmented nucleus (cytosol). Detected RNA particles within the nucleus are marked in red, and those within the cytosol are marked in yellow. Panel (B), shows (from the top left, clockwise) a single-cell characteristic realization of the protein count over time, the joint probability density function~$P({\rm [P], [R_c]})$ from static snapshots of many simulated cells, and the comparison between the true (orange curve) to
the inferred rate~$\mu_{[P]}(\rm [R_c])$, (blue circles, standard deviation
in gray lines). The histogram, in blue shades, represents the observed marginal histogram of variable~$[{\rm R_c]}$. }
\label{fig:BiologicalRealisticModel}
\end{figure}

\subsubsection{Prototypical Examples} 
In these examples, the system has three stochastic variables  ${\vec x} = (x_1, x_2, x_3)$, where $x_1(t)$ evolves as
\begin{equation}
\partial_t x_1(t) = \mu_{1}(x_1,x_2) + \xi_1, 
\label{eq:Proof_of_Concept_Example}
\end{equation}
with $\mu_1(x_1,x_2) = \lambda x_2^n / (x_2^n + K^n) - x_1 /\tau_1$, App.~\ref{App:ProofOfConceptExamples}. The parameters are $\lambda = 80$, $n=2$, $K=40$  and $\tau_1 =1$.  The white noise term, $\xi_1$, is Gaussian distributed with zero mean and $\langle \xi_1(t)\xi_1(t')\rangle = 10^2\delta(t-t')$.  All our simulations were performed by integrating over the Langevin dynamics numerically, using the Euler scheme, see App.~\ref{App:ProofOfConceptExamples}. The coupled dynamical equations for $\vec{x}$ compose network motifs which vary across the examples, leading to  bistable, noise-controlled, and oscillating dynamics.

Figure~\ref{fig:ProofOfConcept} shows that our method successfully infers the shape of the force on $x_1$ as a function of $x_2$, when the self-regularization $\mu_{11}(x_1)=-x_1/\tau_1$ is known.   Decoupled from the effect of $x_1$ on itself, the dependence of the force on $x_1$ on the value of $x_2$, is called $\mu_{12}(x_2)$. The figure shows that our method quantifies the dependence of the force applied on $x_1$ on the state of $x_2$, using only a static observation of $p(x_1,x_2)$. Implementing our method requires no prior knowledge or assumptions made about the mutual influences between the other variables $x_2$ and $x_3$, or about the influence of $x_1$ on those other variables.

 Since we consider forces that do not explicitly depend on time, the inferred $\mu(\vec{x})$ can be used to model and analyze the transient behavior of the system even though we do not record any temporal data. For demonstration, we simulate the systems in App.~\ref{App:ProofOfConceptExamples}, replacing the force between $x_2$ and $x_1$ with the values of $\mu_{12}(x_2)$ that were previously inferred from the data. For all other forces in these simulation we continue to use the analytic value from the models. 

In Fig.~\ref{fig:dynamics_recover}, we present simulation results, comparing between two systems starting from the same initial condition.  Here, in one system  all the forces are known, taken from the model (App~\ref{App:ProofOfConceptExamples}), and in the other system, $\mu_{12}$ inferred from the data.  This figure shows that the system's nontrivial dynamic is recovered, including the transient phase before reaching steady state. 
Recovering transient dynamics from static snapshots of the system at the long-time limit is an indication of the accuracy of our inferred $\mu(\vec{x})$ .  

\begin{figure}[t!]
    \centering
    \includegraphics[width=\columnwidth, trim={100, 0, 80, 0}, clip]{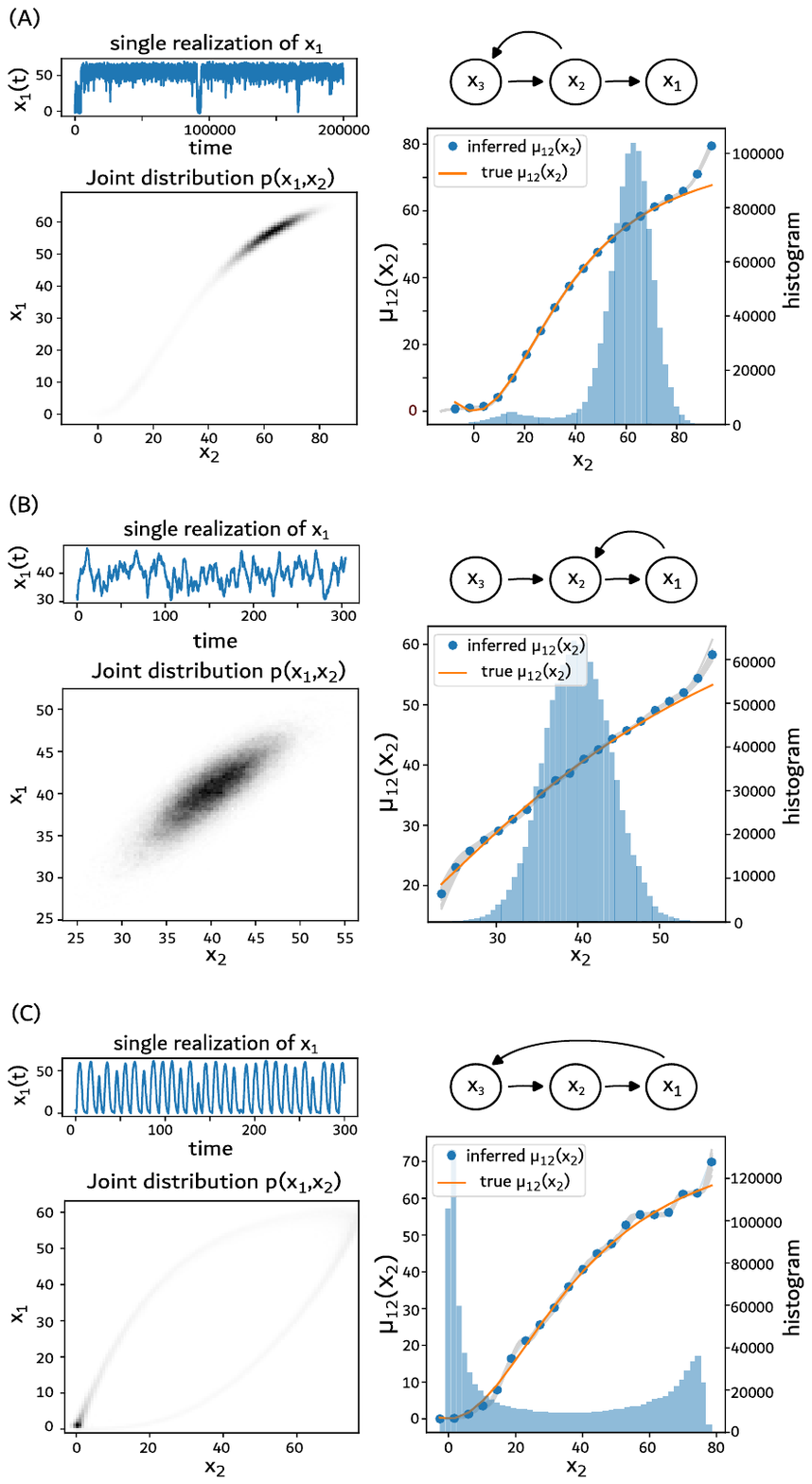}
    \caption{Proof of Concept Examples. In each panel we present the motif topology, a characteristic realization of the variable $x_1(t)$, the joint probability density function $p(x_1,x_2)$, and the comparison between the true force (orange curve) to the inferred force, $\mu_{12}(x_2)$, (blue circles, standard deviation in gray). The histogram, in blue shades, represents the observed marginal histogram of variable $x_1$. The characteristic examples are given as bistable dynamics  (A), noise-controlled (B), and oscillatory dynamics (C), see simulation details in App.~\ref{App:ProofOfConceptExamples}. Our results  display good agreement between the true and inferred $\mu_{12}(x_2)$. }
    \label{fig:ProofOfConcept}
\end{figure}

\begin{figure}
    \centering
   


    \includegraphics[width=\linewidth]{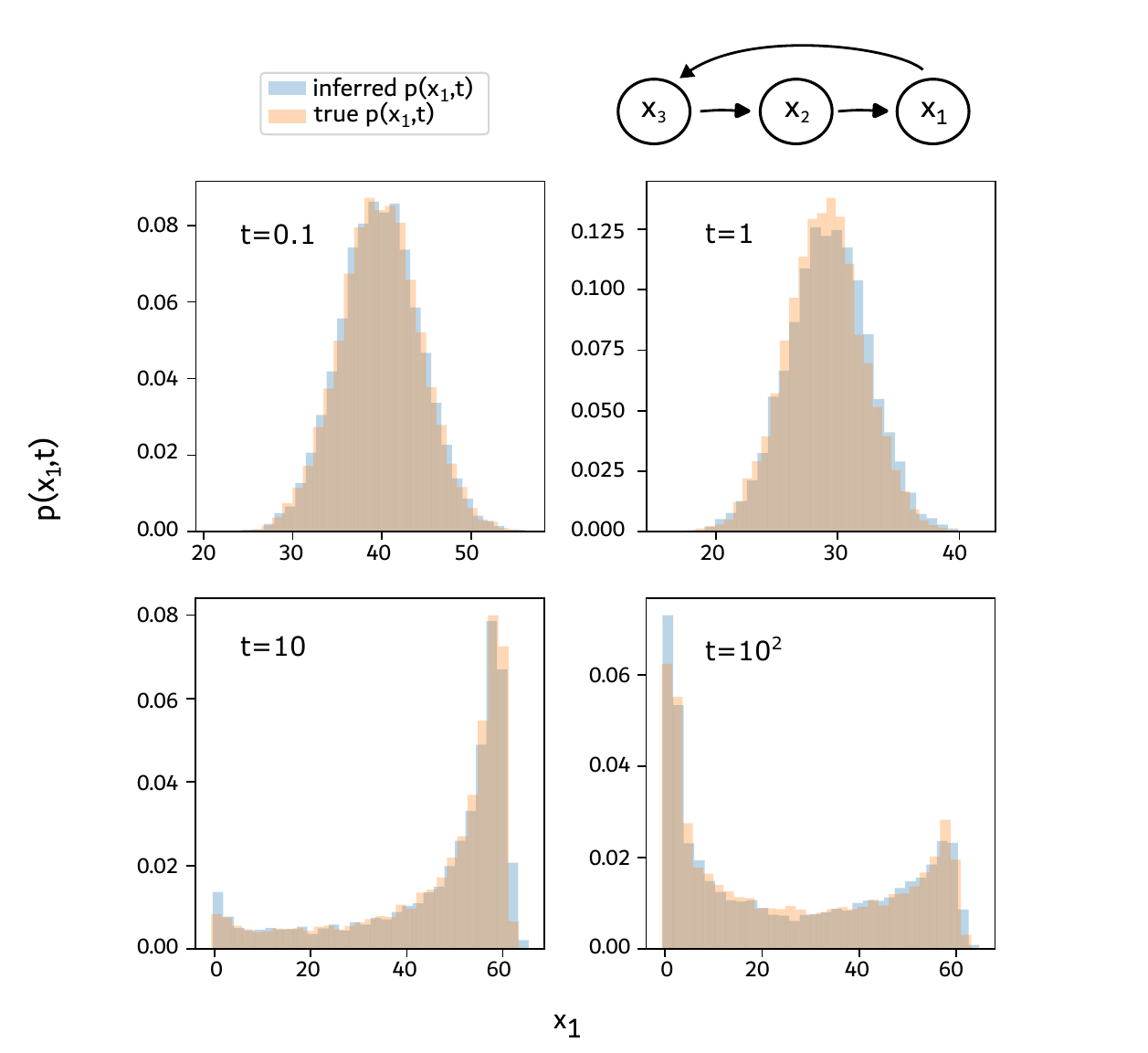}
    \caption{Recovering the dynamics using the inferred force, including transient states.  Comparison between the shape of the marginal probability density, $p(x_1,t)$, obtained at different times from simulation results of the oscillatory dynamics system, App~\ref{App:ProofOfConceptExamples}: Simulation results with the true force appear in orange, and simulation where $\mu_{12}(x_2)$ was quantified from the data appear in blue.  The initial conditions for both simulations were identical. In all these panels, the shape of the probability density in both simulations shows great resemblance, despite the fact that the system has not yet converged to the steady state, indicating that we have successfully inferred the correct shape of the coupling forces. 
    } 
    \label{fig:dynamics_recover}
\end{figure}

\subsection{High Dimensional Examples}

Our method is particularly advantageous when attempting to quantify coupled dynamics in multivariable systems, whereas many other numerical solvers are practically optimized for a relatively small number of variables (less than circa $10$ variables) \cite{sun2014numerical}. Hereby, we demonstrate our method's performance in systems which are high-dimensional in phase space,  using the following two examples. The first example involves 50 fluctuating coupled variables, for which we  quantify at force term in the form of $\mu_{12}(x_2)$. The second example constitutes of 3 variables, where the force on $x_1$, marked as $\mu_{1,23}(x_2, x_3)$ {receives} contribution from the two other variables.
See App.~\ref{App:ThreeComponentsAndMore} and~\ref{App:ProofOfConceptExamples} for details about the simulation and analysis of these systems.}

\begin{figure}
\includegraphics[width=\columnwidth]{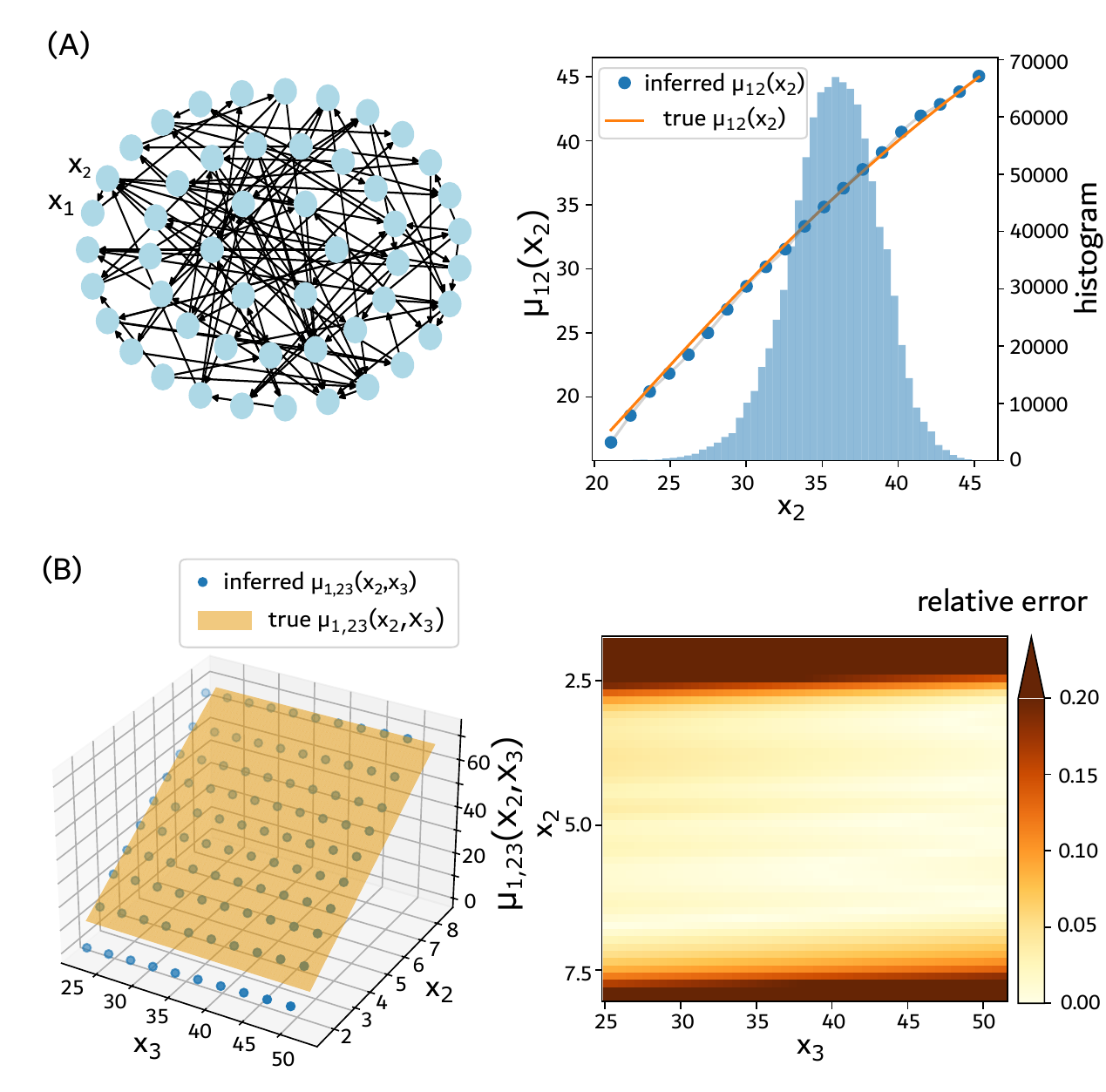}
    \caption{  High-dimensional examples.
    (A)  (left panel) A network consisting of 50 coupled variables is analyzed. (right panel) The comparison involves the actual force (orange curve) and the estimated force $\mu_{12}(x_2)$, shown by blue circles  (gray shade indicates its standard deviation). The observed marginal histogram of variable $x_1$ is seen in blue shade. (B) Quantification of the force when it depends on two variables, $\mu_1(x_2,x_3)$.   The true force is presented as an orange manifold and the force quantified from the data analysis appears in blue circles. The true and inferred forces nicely overlapping the colored manifold (left).  {We can see that the relative error, observed through the color scheme (red means high relative error)},  between the true and inferred forces is mostly smaller than $10\%$, indicating that the  inference of the force is successful (right).  }
\label{fig:HighDimensionalExample_1}
\end{figure}

For our first multivariable example, we consider an Erd\"os–Rényi network with 50 nodes, $\{x_i\}_{i=1}^{50}$, and 100 edges. The dynamics of $x_1(t)$ is the same as Eq.~\eqref{eq:Proof_of_Concept_Example},~App.~\ref{App:ProofOfConceptExamples}. As shown in Fig.~\ref{fig:HighDimensionalExample_1}, the sampled $p(x_1,x_2)$ is sufficient to infer $\mu_{12}(x_2)$, regardless the information about all other variables of the system. 

Up until now, we focused on examples where the applied force on variable $x_1$ depends only on the state of $x_2$. In graph theory terminology, it means that node $x_1$ has only one incoming edge: $x_2 \rightarrow x_1$.  In these cases, only partial measurements of the system state,  i.e. only estimation of $p(x_1,x_2)$ is required. The information on the other variables may remain unrecorded.  However, the suggested method is applicable also where there are multiple incoming edges. For our second multivariable example, we consider 
\begin{equation}
\partial_t x_1(t) = \mu_{1}(x_1,x_2,x_3) + \xi_1 
\label{eq:ManyEdgesExample}
\end{equation}
with $\mu_{1}(x_1,x_2,x_3)= \lambda_{2} x_2 + \lambda_{3} {x_3}^n/({K}^n+{x_3}^n)  - {x_1}/ {\tau_1}$. The parameters are $\lambda_2 = 8$, $\lambda_3=25$, $n=4$, $K=50$ and $\tau_1^{-1}=1$. The simulation results for the inference of $\mu_{1,23}(x_2,x_3)$ are presented in Fig.~\ref{fig:HighDimensionalExample_1}. Note that force applied on $x_1$ depends on states of both variables $(x_2, x_3)$ hence $\mu_{1,23}$ is presented as a two-dimensional manifold. The details are given in Apps.~\ref{App:ThreeComponentsAndMore} and \ref{App:ProofOfConceptExamples}.

\section{Method} 
\label{SecMethod}
\subsection*{Mathematical Background} 
To derive our method, we assume that, in the long-time limit the system can reach a steady state; $p(\vec{x})=\lim_{t\rightarrow\infty}p(\vec{x},t)$. The steady state probability density function satisfies the stationary equation \cite{gardiner1985handbook}:
\begin{equation}
    0 = \sum_{i=1}^N \frac{\partial}{\partial x_i} \left\{ -\mu_i(\vec{x}) \, p(\vec{x})  + \frac{D_{i}}{2}   \frac{\partial}{\partial x_i} p(\vec{x})  \right\}. 
    \label{EqStationaryState}
\end{equation} 
At long enough times, $p(\vec{x},t)\approx p(\vec{x})$, which means that snapshots of $p(\vec{x},t)$ at different times are approximately similar to each other. 

Solving Eq.~\eqref{EqStationaryState} numerically to find the functional shape of the {$\mu_i$s and their dependence on all the variables in $\vec{x}$, using an empirically sampled  $p(\vec{x},t)$, can be challenging, for practical applications. There are two main hurdles~\cite{weinreb2018fundamental}: 
(i) Numerical solvers are inaccurate at high-dimensions~\cite{sun2014numerical}, and (ii) 
The exact multi-dimensional distribution can never be experimentally observed in every point in phase space, but only through a finite number of sampled data points. 
Therefore, to estimate the shape of the $\mu_i$s from a sampled $p(\vec{x},t)$, we leverage a numerical optimization technique \cite{martins2021engineering}, App.~\ref{AppA}. 

Let us isolate the term that we want to evaluate from Eq.~\eqref{EqStationaryState}. Without loss of generality, integrating Eq.~\eqref{EqStationaryState} over all the degrees of freedom besides $i=1$,  with the condition that the probability flux on the boundaries is zero, we get  
\begin{eqnarray}
     \langle  \mu_1(\vec{x}) |x_1\rangle    = \frac{D_{i}}{2}   \frac{1}{p(x_1)}\frac{\partial}{\partial x_1} p(x_1). 
     \label{EqConditionalAverage}
\end{eqnarray}
Here, $\langle (\cdot) |x_1\rangle \equiv \int dx_2 \dots \int dx_N (\cdot) p(x_2, \dots, x_N |x_1)$ means a conditional average for a given $x_1$, and $p(x_1)$ is the marginal probability density.

\subsection*{Solving the Inverse Problem} 
Recall that we are interested in solving Eq.~\eqref{EqStationaryState} using sampled data. But, instead of looking for $p(\vec{x},t)$, we are interested in solving the inverse problem: for a given snapshot of the sampled $p(\vec{x},t)$, we seek the shape of $\mu_i(\vec{x})$. For simplicity, we here present our method for finding $\mu_1(\vec{x})$ for the example where $\mu_1(\vec{x})=\mu_1(x_1,x_2)$ and is independent of all other variables. We consider the case where  $\mu_1(x_1,x_2)$ can be separated as a sum: 
\begin{align}
    \mu_1({x}_1,{x}_2)=\mu_{11}(x_1)-\mu_{12}(x_2). 
    \label{EqDecoupling2D}
\end{align}
Our derivation of the inference method for this example extends also to the case where the dependence on $x_1$ is decoupled as $\mu_1(x_1,x_2) = f(x_1)[\mu_{12}(x_2)-\mu_{11}(x_1)]$, as is commonly assumed, see examples in \cite{levin1976population,hens2019spatiotemporal, osuna2025identification} and references therein.  In the examples in Fig.~\ref{fig:ProofOfConcept}, we chose $f_1=1$, thus the following derivation corresponds to that choice. The more general case is  discussed in App.~\ref{AppA}.

The probability density function is sampled at various data points $p(x_1^i,x_2^j)$, for every $x_1^i \in (x_1^1, x_1^2,\dots, x_1^{M_1})$ and $x_2^j \in (x_2^1, x_2^2,\dots, x_2^{M_2})$. Here, the lower index represents the variable index $1$ or $2$, and the upper index represents the sequential order of the points. The sequence spacing in $\{x_1^{i}\}$ and $\{x_2^{j}\}$  is  $\Delta x_1$ and $\Delta x_2$, respectively. This means that the sampled values  $p(x_1^i,x_2^j)$ are terms in a  $M_1\times M
_2$ matrix, $\hat{{ \rm P}}$:   
{
\begin{align}
    \hat{{\rm P}}&=\begin{pmatrix}
p(x_1^1,x_2^1),& \cdots, & p(x_1^{1}, x_2^{M_2})\\
p(x_1^2,x_2^1),& \cdots, & p(x_1^{2},x_2^{M_2})
\\
\vdots& \vdots & \vdots\\
p(x_1^{M_1},x_2^{1}),& \cdots, & p(x_1^{M_1},x_2^{M_2}) 
\end{pmatrix}\Delta x_2,
\label{EqMatrix2D} 
\end{align} 
} 
where  each value in the matrix $p(x_1^i, x_2^j)$ is subject to sampling noise, as  discussed   in Ref. \cite{leibovich2025determining}. 
Using Eqs.~\eqref{EqConditionalAverage}-\eqref{EqMatrix2D}, we obtain a set of $M_1$ equations:  
{ \begin{widetext}
    \begin{align}
      & \mu_{12}(x_2^1) p\left(x_1^1,x_2^1\right) \Delta x_2  + \mathinner{\dots} + \mu_{12}(x_2^{M_2}) p\left(x_1^1,x_2^{M_2}\right) \Delta x_2 = \mu_{11}\left(x_1^1\right)p(x_1) + \frac{D_{1}}{2}   \frac{\Delta  p(x_1)}{\Delta x_1}|_{x_1^1} \nonumber \\
      & \mu_{12}\left(x_2^1\right) p\left(x_1^2,x_2^1\right) \Delta x_2  + \mathinner{\dots}  \mu_{12}\left(x_2^{M_2}\right) p\left(x_1^2 ,x_2^{M_2}\right) \Delta x_2 = \mu_{11}\left(x_1^2\right) p\left(x_1^2\right) + \frac{D_{1}}{2}   \frac{\Delta  p(x_1)}{\Delta x_1}|_{x_1^2} \nonumber \\
      & \vdots \nonumber \\
      & \mu_{12}(x_2^{1}) p(x_1^{M_1},x_2^1) \Delta x_2  + \dots +\mu_{12}(x_2^{M_2}) p(x_2^{M_1},x_2^{M_2}) \Delta x_2 = \mu_{11}(x_1^{M_1}) p(x_1^{M_1}) + \frac{D_{1}}{2}   \frac{\Delta p(x_1)}{\Delta x_1}|_{x_1^{M_1}},
      \label{EqEqsSet}
    \end{align}
\end{widetext}} where $\Delta p(x_1) |_{x_i^i} = p(x_1^{i})-p(x_1^{i-1})$, see App.~\ref{AppA}. 
The Eq. set~\eqref{EqEqsSet} can be compacted into a single linear matrix equation: 
\begin{align}
    \hat{ \rm P}\cdot \vec{\mu}_{12} =\vec{b}.  
    \label{EqLinearMatrixEquation} 
\end{align} 
Each term, $j\in[1\dots M_2]$, in $\vec{\mu}_{12}$ is given by the value of $\mu_{12}(x_2^j)$. Similarly, each term $i\in[1\dots M_1]$, in vector $\vec{b}$ is $\left[\mu_{11}(x_1^i)p(x_1^i) + \frac{D_1}{2}\frac{\Delta p(x_1)}{\Delta x_1}|_{x_1^i}\right]$. Recovering the value of $\vec{\mu}_{12}$ in every $x_2^j$,  using Eq.~\eqref{EqLinearMatrixEquation}, means inferring the function $\mu_{12}$ from the data. In other words, this means quantifying the coupling between the system's dynamical variables, $x_1$ and $x_2$, from a snapshot of the sampled $p(\vec{x})$, without the need for modeling.   

\subsection*{Optimal Solution from Numerical Minimization.} 
If we know the {\em exact} value of $p(\vec{x})$ everywhere in phase space, we can infer the shape of $\mu_{12}(x_2)$ exactly for any value of $x_2$. Because measured distributions always have noise,  emerging from finite sample size and discretizations, for practical applications one can solve Eq.~\eqref{EqLinearMatrixEquation} using a numerical optimization technique \textit{resembling numerical regression}, App.~\ref{AppA}. 
We seek to minimize a \textit{loss function}, $|\hat{P}\cdot\vec{\mu}_{12} - \vec{b}|^2$, with an additional regularization term $\lambda |\hat{R}\vec{\mu}_{12}|^2$ to prevent overfitting and to penalize non-realistic solutions of $\vec{\mu}_{12}$ (see the appendix).  Now, the values of $\mu_{12}(x_2^j)$ for every $j\in[1,2\dots M_2]$ are obtained from the sampled probability density by minimizing the regularized loss function \begin{equation}
    \mu_{12} = {\rm arg}\min_{\mu_{i2}} \{ |\hat{\rm P}\cdot\vec{\mu}_{12} - \vec{b}|^2 + \lambda |\hat{R}\vec{\mu}_{12}|^2 \}. 
    \label{EqGeneralMinimization2D}
\end{equation}
To reduce overfitting in the numerical minimization, $\mu_{12}(x_2)$ may be subjected to constraints emerging from real-world scenarios. For the examples presented in Figs. (\ref{fig:BiologicalRealisticModel}-\ref{fig:HighDimensionalExample_1}), we used $\lambda = 10^{-3}$, and the shape of $\hat{R}$ chosen in order to help recover a $\vec{\mu}_{12}(x_2)$ which is smooth, App.~\ref{AppA}.  In addition, we subject the solution $\mu_{12}(x_2)$ to be a  monotonous and non-negative function to mimic biological systems, e.g. \cite{wittenstein2022quantifying}. The minimized solution is obtained numerically, see App.~\ref{AppA}.

\section{Discussion}
We have derived a method for quantifying complex coupled dynamics in multivariable systems from the stationary state of their probability density function. Our results show that this method is useful, for example, for the analysis of probability densities obtained from experiments using spectroscopic imaging.  We demonstrated that for the variable under investigation $x_1$, if its dynamics is externally influenced by a variables set, ${\vec{E}_1^-}$, we can quantify the full shape of the force, $\mu_1({\vec{E}_1^-})$, using only partial information about the probability density $p(x_1,{\vec{E}_1^-})$. Our method allows the information from all other confounders within the system to remain undetermined.    The advantage of sampling from a steady state lies in its ability to record data without requiring temporal ordering. Eliminating the requirement for temporal ordering enables data collection across various sampling intervals, or aggregated from multiple experiments, provided they are conducted under the same conditions. Consequently, the assumption of stationarity facilitates the analysis for various systems~\cite{friedman2006linking, taniguchi2010quantifying, larvie2016stable}. 

 We quantify coupling forces between  various variables by solving an optimization problem, Eqs. (\ref{EqMainResults5}) and (\ref{EqGeneralMinimization2D}). Our quantification approach is also applicable to rotational force fields, complementing previously  published works~\cite{weinreb2018fundamental}. Notably, it reveals the force dependence without being limited by specific functional forms or mechanistic models \cite{brunton2016discovering, timme2007revealing, timme2014revealing}. Our method enables to quantify also other dynamics, such as first passage times, see App.~\ref{App:More_sim_Results}.

 Deep learning methods could support our method, in future research. Previously published techniques use artificial neural networks to study the direction in pairwise interaction in complex network, e.g. \cite{goudet2018learning}. This approach can be extended to quantify the interaction forces. Importantly, deep learning approaches methods require large dataset for training and validating stages, and strong computational resources. However, these constraints may be improved with future technological advancements. 
 
\newpage

\appendix

\begin{widetext}
\section{Additional details on the derivation of Eq.~\eqref{EqEqsSet} and solution to  Eq.~\eqref{EqGeneralMinimization2D}}
\label{AppA}
{\textbf{Details on the derivation of Eq.~\eqref{EqEqsSet}
.} 
Using Eq.~\eqref{EqDecoupling2D}, Eq.~\eqref{EqConditionalAverage} can be written as
\begin{eqnarray}
      \int_{-\infty}^\infty dx_2 \mu_{12}(x_2) p({x}_1,x_2) = \mu_{11}(x_1) p(x_1) + \frac{D_{1}}{2}   \frac{d p(x_1)}{dx_1}.
\end{eqnarray}
Discretizing the integral and the derivatives, for every possible value of $x_1$, we get 
\begin{eqnarray}
     \sum_{j=-\infty}^{\infty}  \mu_{12}(x_2^j) p(x_1,x_2^j) \Delta x_2 = \mu_{11}(x_1) p(x_1) + \frac{D_{1}}{2}   \frac{d p(x_1)}{d x_1},  
     \label{EqAppAInfiniteSum}
\end{eqnarray} 
where $\Delta x_2\ll1$. 
{Since $p(x_1,x_2)$, Eq.~\eqref{EqMatrix2D}, is sampled from experimental measurements or simulation results} there is only a finite region; $\{x_1^1\dots x_1^{M_1}\}$ by $\{x_2^1\dots x_2^{M_2}\}$, where $p(x_1,x_2)\neq 0$. Therefore, instead of an infinite sum in Eq.~\eqref{EqAppAInfiniteSum}, for every value $x_1^i$, where $i=1,2,\dots M_1$, we have  
\begin{eqnarray}
     \sum_{j=1}^{M_2}  \mu_{12}(x_2^j) p(x_1^i,x_2^j) \Delta x_2 = \mu_{11}(x_1^i) p(x_1^i) + \frac{D_{1}}{2}   \frac{\Delta p(x_1)}{\Delta x_1}|_{x_1^i}. 
     \label{EqAppAFiniteSum1}
\end{eqnarray} 
From Eq.~\eqref{EqAppAFiniteSum1}, for each term $x_1^i$ ($i=1,2,\dots, M_1$) we get one line in the equation set, Eq.~\eqref{EqEqsSet}. 

\textbf{Solving a minimization problem. } The set of equations, Eq.~\eqref{EqEqsSet}, for every $x_1^i$, can be represented as  $\hat{P}\vec{\mu}_{12}=\vec{b}$, Eq.~\eqref{EqLinearMatrixEquation}. The quantification procedure of $\mu_{12}(x_2^j)$ for the values of $x_2\in\{x_2^1,\dots x_2^{M_2}\}$ means to find which  set of $\mu_{12}(x_2^j)$ minimizes} [{LHS (Eq.\eqref{EqAppAFiniteSum1}) - RHS(Eq.\eqref{EqAppAFiniteSum1})}]$^2$ for every observed $p(x_1^i,x_2^j)$. In other words, one aims to find the set $\{\mu_{12}(x_2^j)\}_{j=1}^{M_2}$ that minimizes $|\hat{P}\vec{\mu}_{12}-\vec{b}|^2$. To reduce the chance of overfitting or obtaining non-realistic solutions of $\mu_{12}$ emerging from the sampling noise of $p(x_1^i, x_2^j)$ propagating to $\mu_{12}$, we add an additional regularization term such that
$
    \mu_{12} = {\rm arg}\min_{\mu_{12}} \{ |\hat{\rm P}\cdot\vec{\mu}_{12} - \vec{b}|^2 + \lambda |\hat{R}\vec{\mu}_{12}|^2 \}$
as is given in Eq.~\eqref{EqGeneralMinimization2D}. Since the examples examined in this manuscript are inspired by biochemical interaction networks, we regularized the solution $\mu_{12}(x_2)$ to be non-zero, monotonous smooth function. Particularly for monotonously increase function we solve
\begin{equation}
    \mu_{12} = {\rm arg}\min_{\mu_{i2}} \{ |\hat{\rm P}\cdot\vec{\mu}_{12} - \vec{b}|^2 + \lambda |\hat{R}\vec{\mu}_{12}|^2 \} {\rm \ \ subjected \ to \ \ } \forall j: \mu_{12}(x_2^j) \geq 0 {\rm \ \ and \ \ } \forall j\neq 1 : \mu_{12}(x_2^j) > \mu_{12}(x_2^{j-1}). 
    \label{EqA4}
\end{equation}
Here, each term in the matrix $\hat{R}$ is $R_{i,j}=\delta_{i,j} - 2\delta_{i,j+1} + \delta_{i,j+2}$. This choice of regulation promotes   smoothness of $\mu_{12}(x_2)$ as follows:  Every row in $\hat{R}\mu_{12}(x_2)$ is given by $ \mu_{12}(x_2^{j+2}) -2 \mu_{12}(x_2^{j+1})+\mu_{12}(x_2^j)$, which is the numerical 2nd order derivative of $\mu_{12}(x_2)$. In Eq.~\eqref{EqA4} one minimizes this second derivative of the coupling force alongside the term $|\hat{P}\vec{\mu}_{12}-\vec{b}|^2$, derived from the stationary Fokker-Planck equation. The minimization problem is numerically solved using the  Python library `cvxopt' \cite{CVXOPT}.
\\

\textbf{A note about a more generalized decoupling form of $\mu_1(\vec{x})$. }
As mentioned in the main text, a more generalized form of the decoupling in Eq.~\eqref{EqDecoupling2D} is expressed as $\mu_1(x_1,x_2) = f(x_1)[\mu_{12}(x_2)-\mu_{11}(x_1)]$. Here, the shapes of $\mu_{11}(x_1)$ and $f(x_1)$ are assumed to be known and we seek $\mu_{12}(x_2)$. Starting from Eq.~\eqref{EqConditionalAverage},  substituting the general decoupling form of the force, 
we obtain
\begin{equation}
    f(x_1^i) \sum_{j=1}^{M_2} \Delta x_2 \mu_{12}(x_2^j) p(x_1^i,x_2^j) = f(x_1^i)\mu_{11}(x_1^i)p(x_1^i) + \frac{D_i}{2}\frac{\Delta  p(x_1)}{\Delta x_1}.
\end{equation}
In a similar manner to the method presented when $f=1$, we infer $\mu_{12}(x_2)$ by finding 
\begin{equation}
    \mu_{12} = {\rm arg}\min_{\mu_{12}} \{ |\hat{\rm P}^*\cdot\vec{\mu}_{12} - \vec{b}^*|^2 + \lambda |\hat{R}\vec{\mu}_{12}|^2 \}
\end{equation}
replacing ${P}_{ij} = p(x_1^i, x_2^j) \Delta x_2$ with ${P}^*_{ij} = f(x_1^i)p(x_1^i, x_2^j)\Delta x_2$, and $b_i = \mu_{11}(x_1^i)p(x_1^i) + D_i \Delta p(x_1^i)/(2\Delta x_1)$. The regularization matrix $\hat{R}$ and the solution constrains are chosen corresponding the nature of the system under investigation.  The rest of the approach remains unchanged.

 \section{Extension to systems with three components and more} 
 \label{App:ThreeComponentsAndMore}
In here, we assume that the applied force on $x_1$ depends on two variables; $x_2$ and $x_3$, and we want to compute $\mu_{1, 23}(x_2,x_3)$ from $P(x_1,x_2,x_3)$. In principle, the method is the same as described above. The key in this case is that the matrix $\mu(x_2^j,x_3^k)$ (of size $M_2\times M_3$) can be ``rolled-up''   into a vector $\mu_{1,23}(y)$, namely  
\begin{equation*}
\underbrace{(x_2,x_3)}_{{\rm size \ } M_2 \times M_3} \longrightarrow \underbrace{(y)}_{{\rm size\ } 1 \times M_2 M_3}
\end{equation*}
with
\begin{equation*}
 {y}\equiv \{(x_2^1,x_3^1),\dots,(x_2^{M_2},x_3^1),(x_2^1,x_3^2),\dots,(x_2^{M_2},x_3^2),\dots, \dots,  (x_2^{M_2},x_3^{M_3}) \}.
\end{equation*}
With this vector, $y$, we can use a similar approach to that given in the main text and App.~\ref{AppA} to find $\mu_{1,23}(y)$, then reshape the latter vector back into a matrix $\mu_{1,23}(x_2,x_3)$. One small change needs to be applied to the minimization problem:
\begin{eqnarray}
   && \mu_{12} = {\rm arg}\min_{\mu_{i2}} \{ |\hat{\rm P}\cdot\vec{\mu}_{12} - \vec{b}|^2 + \lambda |\hat{R}\vec{\mu}_{12}|^2 \}   {\rm \ \ subjected \ to \ \ }  [\vec{\mu}_{12} {\rm \ positive \ \&\ monotonous}].
   \nonumber
\end{eqnarray}
\textcolor{black}{The constrains on  $\vec{\mu}_{12}$ correspond to constraints $\vec{y}$
. 
We condition $\mu_{12}(x_2,x_3)$ to be monotonous in each direction $x_2$ and $x_3$ individually. For a non-decreasing function in both directions, we force the conditions (i) $\mu_{12}(x_2^{i+1},x_3^j)\geq \mu_{12}(x_2^i,x_3^j)$ and  (ii)  $\mu_{12}(x_2^{i},x_3^{j+1})\geq \mu_{12}(x_2^i,x_3^j)$, for every $(i,j)$. The non-decreasing behavior is expressed by the matrix $\hat{d}_1\equiv -\delta_{ij}+ \delta_{i+1, j}$. Then, we define $\hat{A}_{\rm rows} \equiv \hat{d_1} \otimes \hat{I}$, and $\hat{A}_{\rm cols} \equiv  \hat{I} \otimes \hat{d_1} $. Here, $\delta_{ij}$ is the Kronecker delta function, and $\otimes$ is the Kronecker product. The matrix $\hat{A}$, setting the constraint $\hat{A}\cdot \vec{y}\geq 0$, is defined by concatenating $\hat{A}_{\rm rows}$ and $\hat{A}_{\rm cols}$. The matrix $\hat{I}$ is the unit (identity) matrix.} 
 A similar definition takes place for the 2nd order derivative with $\vec{d}_2 = -2\delta_{i,j}+\delta_{i+1,j}+\delta_{i-1,j}$. 

\section{Simulation Details}
\label{App:ProofOfConceptExamples}

 {In our examples, whose results are presented in Fig.~\ref{fig:BiologicalRealisticModel} and Fig.~\ref{fig:ProofOfConcept}, we use a simulation of a model of gene regulation~\cite{aguilera2025methods}, and three simulations of prototypical interaction networks}. We simulated the complex dynamics using the Euler method \cite{butcher2016numerical},  up to arbitrary long times, $t$, until we observed that the  simulated system has reached steady state. We then sampled from the numerically observed stationary state $n$ data points $\{(x_1^i, x_2^i)\}_{i=1}^n$. Note that information about the dynamics of other parts of the system is not recorded, demonstrating that our inference methods does not require it. From the set of $n$ data points we numerically infer the joint probability density function $p(x_1,x_2)$ using multi-dimensional normalized histogram of the data.

\begin{itemize}
    \item {\em Gene-Expression Simulation} 
     The dynamical equations and parameters of this example are as follows: 
\begin{eqnarray}
\nonumber
    \partial_t [R_n] &=& k_r-k_t [R_n]-\gamma_r [R_n] + \xi_n\\ 
     \partial_t [R_c] &=&  k_t [R_n]   - \gamma_r [R_c]+\xi_c \\
    \partial_t [P] &=&   k_p [R_c]-\gamma_p [P]+\xi_p,  
    \nonumber 
    \label{EqGeneRegulationSimulation}
\end{eqnarray}
where $[R_n],[R_c]$ and $[P]$ denote the concentrations of the mRNA nuclear, mRNA cytoplasm and protein concentrations (respectively). We note that we use the assumption provided in  \cite{aguilera2025methods} - that the gene is always active, and the concentration of the gene in the system remains constant. Therefore, [G] is not explicitly simulated, and its concentration is incorporated into the constant $k_r$.   The parameter $k_r$ represents the {\em constitutive} production rate of nuclear mRNA, $k_t$ is the mRNA transport rate to cytoplasm, and $k_p$ is the production rate of proteins. The constants $\gamma_r$ and $\gamma_p$ are the decay rates of the mRNA (both nuclear and  cytoplasm) and protein respectively.  The values of the parameters used in our simulations are: $k_r=3, k_t=0.083, k_p =0.9, \gamma_r=0.08, \gamma_p=0.45$ as provided in \cite{aguilera2025methods}. All the rates are provided with units of $[{\rm min}]^{-1}$, and the variables $[{\rm R_n}], [{\rm R_c}]$ and $[{\rm P}]$ are in dimensionless units~\cite{aguilera2025methods}.   The additive noise terms $\xi_n, \xi_c$ and $\xi_p$, which can be emerge in experimental data for example from the sampling noise, were obtained in our simulations independently from each-other, from a Gaussian distribution with zero mean and variance $\sigma=10$, for all the variables. \\
\end{itemize} 

For the three prototypical interaction models, we considered complex systems described by coupled stochastic differential equations as follows. Each of the following example is named by the obtained characteristic dynamics: \begin{itemize}
    \item 
    {\em Example 1 - Bistable Dynamics}
\begin{eqnarray}
\nonumber
    \partial_t x_3 &=& \lambda_3 + \lambda_{3,1} \frac{{x_1}^{n_3}}{{K_3}^{n_2}+{x_1}^{n_3}} - {x_2}/ {\tau_3}  +\xi_3\\ 
     \partial_t x_2 &=&  \lambda_{2,3} x_3   - {x_2}/ {\tau_2}  +\xi_2 \\
    \partial_t x_1 &=&   \lambda_{1,2} \frac{{x_2}^{n_1}}{{K_1}^{n_1}+{x_2}^{n_1}}  - {x_1}/ {\tau_1}  +\xi_1  
    \nonumber
\end{eqnarray}
with the parameters $\lambda_3 = 0.45, \lambda_{3,1} = 1.5, n_3 = 6, K_3 = 37, \tau_3^{-1} = 0.03$, $\lambda_{2,3} = 0.03, \tau_2^{-1} = 0.03 $, $\lambda_{1,2} = 80,  n_1 = 2, K_1 = 40, \tau_1^{-1} = 1 $. The additive noise $\xi_i$ is Gaussian distributed with zero mean and scale $\sigma_i = 10$ for all $i$.
\item 
{\em Example 2 - Noise Controlled}
\begin{eqnarray}
\nonumber
    \partial_t x_3 &=& \lambda_3  - {x_3}/ {\tau_3}  +\xi_3\\ 
     \partial_t x_2 &=&  \lambda_{2,3} x_3 + \lambda_{2,1} \frac{{x_1}^{n_2}}{{K_2}^{n_2}+{x_1}^{n_2}}  - {x_2}/ {\tau_2}  +\xi_2 \\
    \partial_t x_1 &=&   \lambda_{1,2} \frac{{x_2}^{n_1}}{{K_1}^{n_1}+{x_2}^{n_1}}  - {x_1}/ {\tau_1}  +\xi_1
    \nonumber
\end{eqnarray}
with the parameters $\lambda_3 = 5, \tau_3^{-1} = 1$, $\lambda_{2,3} = 8, \lambda_{2,1}=25, n_2=4, K_2 = 50,  \tau_2^{-1} = 1 $, $\lambda_{1,2} = 80,  n_1 = 2, K_1 = 40, \tau_1^{-1} = 1 $. The additive noise $\xi_i$ is Gaussian distributed with zero mean and scale $\sigma_i = 10$ for all $i$.
\item 
{\em Example 3 - Oscillatory Dynamics}
\begin{eqnarray}
\nonumber
    \partial_t x_3 &=& \lambda_{3,1} \frac{{K_3}^{n_3}}{{K_3}^{n_3}+{x_1}^{n_3}} - {x_3}/ {\tau_3}  +\xi_3\\ 
     \partial_t x_2 &=&   \lambda_{2,3} \frac{{x_3}^{n_2}}{{K_2}^{n_2}+{x_3}^{n_2}}  - {x_2}/ {\tau_2}  +\xi_2 \\
    \partial_t x_1 &=&   \lambda_{1,2} \frac{{x_2}^{n_1}}{{K_1}^{n_1}+{x_2}^{n_1}}  - {x_1}/ {\tau_1}  +\xi_1
    \nonumber
\end{eqnarray} 
with the parameters $\lambda_{3,1} = 50000, K_3 = 0.1, n_3 = 10, \tau_3^{-1} = 1$, $\lambda_{2,3} = 80,  n_2=2, K_2 = 40,  \tau_2^{-1} = 1 $, $\lambda_{1,2} = 80,  n_1 = 2, K_1 = 40, \tau_1^{-1} = 1 $. The additive noise $\xi_i$ is Gaussian distributed with zero mean and scale $\sigma_i = 10$ for all $i$.        
    \item  
{\em Example 4 - Network with 50 Coupled Variables}
The dynamics of the process is described by
\begin{eqnarray}
\nonumber
    \partial_t x_i &=& 1 - x_i/\tau_i + \sum_{j=2}^{50} G_{ij} \cdot \frac{\lambda _i x_{j}^{n_j}}{K_j^{n_j}+x_j^{n_j}} +\xi_i  {\rm \ for\ } i\neq 1,   \\
    \partial_t x_{1} &=& \frac{\lambda_1 x_{2}^{n_1}}{K_1^{n_1}+x_{2}^{n_1}} -  x_{1} + \xi_1.   
\end{eqnarray}
with $\forall j \neq  : \lambda_j = 100,  \tau_j^{-1}=1.3, K_j = 100, n_j = 1$. For the $x_1$: $\lambda_1 = 80$, $n_1=2$ and $K_1=40$. 
Here, $\hat{G}$ is a random Erdős–Rényi network with 50 nodes and 100 edges. 

\item 
{\em Example 5 - Quantified Influence from Two Incoming Edges}

We use the same system as provided in `Example 2 - Noise Control' above. However, instead of concentering on the influences on $x_1$, we look at the the variable $x_2$. The latter has two incoming edges - from $x_1$ and $x_3$.  

\end{itemize}

\section{Additional Simulation Results - Temporal Features Recovery}
\label{App:More_sim_Results}

In the Discussion, we state that our method enables to quantify additional observables, beyond the coupling forces. 
For the Oscillatory dynamics, App.~\ref{App:ProofOfConceptExamples}, using the simulation results from Fig.~\ref{fig:ProofOfConcept}, in the main text, Fig.~\ref{fig:temp_recovery_additional_data} shows our successful recovery of the time dependent mean $\langle x_1 (t)\rangle$, the standard deviation ${\rm std}(x_1)(t)$, and the occupation time in an arbitrarily interval ($35<x<45$).

\begin{figure}
    \centering
    
\begin{flushleft}
\hspace{5cm }(A)
\end{flushleft}
    
    \includegraphics[width=0.5\columnwidth]{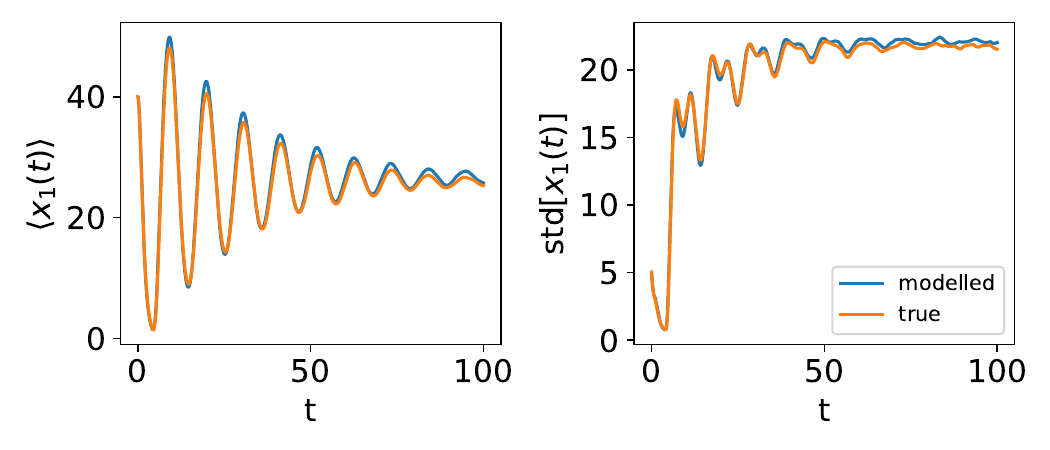}

\begin{flushleft}
\hspace{5cm }(B)
\end{flushleft}
    \includegraphics[width=0.4\columnwidth]{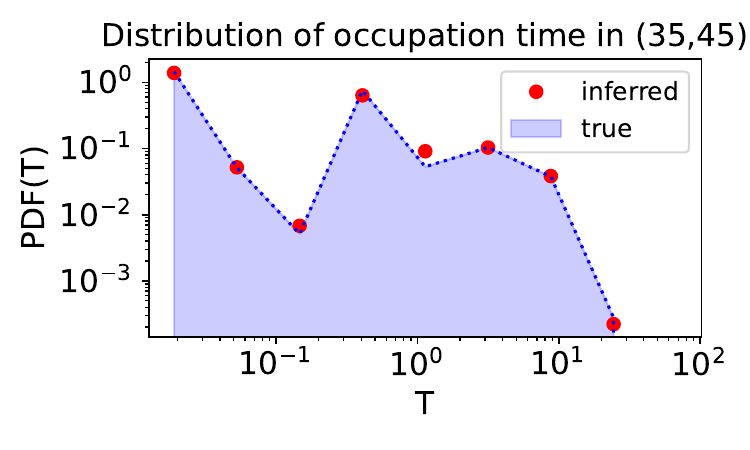}
    \caption{Dynamical Features recovered. The simulation results are given for the oscillatory dynamic example. Panel (A) shows the mean (left) and standard deviation (right) of $x_1$; blue: results of a simulation where forces are known from the model, App.~\ref{App:ProofOfConceptExamples}, orange: simulation results where the coupling force was inferred from data. Panel (B) shows the probability density function  PDF(T) of the sojourn times (T) within the interval $(35,45)$. Here the results obtained from the true dynamics are represented by a blue shade, and results obtained when $\mu_{12}$ inferred from data is given by red circles.     }
\label{fig:temp_recovery_additional_data}
\end{figure}

\newpage

\end{widetext}


\bibliography{apssamp}

\end{document}